\documentclass[11pt, letterpaper]{article}

\usepackage{amsmath}
\usepackage{amssymb}
\usepackage{amsthm}
\usepackage{graphicx}
\usepackage{hyperref}
\usepackage{url}
\usepackage{algorithm}
\usepackage{algorithmic}
\usepackage{caption}
\usepackage{subcaption}
\usepackage{booktabs}
\usepackage{authblk}
\usepackage{appendix}
\usepackage[utf8]{inputenc}
\usepackage[T1]{fontenc}
\usepackage[english]{babel}
\usepackage{geometry}
\usepackage{xcolor}
\usepackage{fancyhdr}
\usepackage{microtype}
\usepackage{parskip}
\usepackage{tabularx}
\usepackage{enumitem}

\hypersetup{
    colorlinks=true,
    linkcolor=blue,
    filecolor=magenta,      
    urlcolor=cyan,
    citecolor=red,
}

\definecolor{headercolor}{RGB}{0, 50, 100}
\title{Artificial General Intelligence for Radiation Oncology}

\author[1]{Chenbin Liu, PhD}
\author[2]{Zhengliang Liu, MS}
\author[3]{Jason Holmes, PhD}
\author[4]{Lu Zhang, MS}
\author[3]{Lian Zhang, PhD}
\author[3]{Yuzhen Ding, PhD}
\author[2]{Peng Shu, MS}
\author[2]{Zihao Wu, MS}
\author[2]{Haixing Dai, PhD}
\author[2]{Yiwei Li, MS}
\author[5,6,7]{Dinggang Shen, PhD}
\author[2]{Ninghao Liu, PhD}
\author[8]{Quanzheng Li, PhD}
\author[8]{Xiang Li, PhD}
\author[4]{Dajiang Zhu, PhD}
\author[2]{Tianming Liu, PhD}
\author[3]{Wei Liu, PhD}

\affil[1]{Department of Radiation Oncology, National Cancer Center\//National Clinical Research Center for Cancer\//Cancer Hospital \& Shenzhen Hospital, Chinese Academy of Medical Sciences and Peking Union Medical College, Shenzhen, Guangdong, China.}
\affil[2]{School of Computing, University of Georgia}
\affil[3]{Department of Radiation Oncology, Mayo Clinic}
\affil[4]{Department of Computer Science and Engineering, The University of Texas at Arlington}
\affil[5]{School of Biomedical Engineering, ShanghaiTech University}
\affil[6]{Shanghai United Imaging Intelligence Co., Ltd.}
\affil[7]{Shanghai Clinical Research and Trial Center}
\affil[8]{Department of Radiology, Massachusetts General Hospital and Harvard Medical School}

\date{}

\begin{document}

\maketitle

\begin{abstract}
The emergence of artificial general intelligence (AGI) is transforming radiation oncology. As prominent vanguards of AGI, large language models (LLMs) such as GPT-4 and PaLM 2 can process extensive texts and large vision models (LVMs) such as the Segment Anything Model (SAM) can process extensive imaging data to enhance the efficiency and precision of radiation therapy. This paper explores full-spectrum applications of AGI across radiation oncology including initial consultation, simulation, treatment planning, treatment delivery, treatment verification, and patient follow-up. The fusion of vision data with LLMs also creates powerful multimodal models that elucidate nuanced clinical patterns. Together, AGI promises to catalyze a shift towards data-driven, personalized radiation therapy. However, these models should complement human expertise and care. This paper provides an overview of how AGI can transform radiation oncology to elevate the standard of patient care in radiation oncology, with the key insight being AGI's ability to exploit multimodal clinical data at scale.
\end{abstract}

\section{Introduction}

An estimated 600,000 people in the United States die from cancer every year. Beyond surgery and chemotherapy (now augmented by immunotherapy), radiotherapy has proven as a standard and effective treatment option for nearly 50-70 percent of cancer patients. There exist diverse modalities for the delivery in radiotherapy: (1) brachytherapy involves the surgical implantation of radioactive sources into the patient to kill tumors\cite{nag1999american}; (2) early external beam radiotherapy utilizes strong radioactive sources like Cobalt60 positioned at a distance from patients with collimators used to shape and direct beams\cite{lott1958cobalt}; (3) contemporary external beam radiotherapy utilizes medical linear accelerator to generate high-energy electron beams or photon beams via bremsstrahlung interactions, optimized for tumor targeting through treatment planning systems\cite{siebers2008quantification}; (4) particle therapy with protons or heavier ions produces conformal dose distributions with reduced exit doses, demonstrating efficacy in select cases \cite{mohan2017proton, schild2014proton, deng2021critical, liu2019system}. 

Radiotherapy treatment involves six basic stages: initial consultation, simulation, treatment planning, treatment delivery, treatment verification, and patient follow-up. The initial consultation includes a radiation oncologist reviewing the patient’s medical history including demographics, operative notes, pathology reports, radiology reports, lab results, discharge, and consults notes to determine the appropriateness of radiotherapy \cite{chan2022artificial}. Simulation precisely localizes the tumor using CT/MR imaging and customized immobilization devices to ensure reproducibility between fractions. In treatment planning, medical professionals delineate target and organ-at-risk volumes on simulation images\cite{liu2015impact}. Medical dosimetrists and physicists design individualized treatment plans balancing tumor control and normal tissue toxicity\cite{webb1993model, zaghian2014automatic, zaghian2017comparison}. Treatment verification involves the evaluation of dosimetric and geometric accuracy for radiation safety\cite{younkin2018multiple, younkin2018efficient}. Qualified radiation therapists perform image-guided radiation delivery as prescribed. Periodic patient follow-up monitors the progress and addresses any side effects. Longitudinal surveillance further guides the management of potential recurrence or residual disease\cite{nathan2020early}\cite{yu2022cardiopulmonary}.  With the emergence of deep learning algorithms, radiotherapy is undergoing substantial transformation. Artificial intelligence (AI) achieves human-level accuracy in the auto-segmentation of organs-at-risk and tumor volumes from CT/MR images, which saves clinicians' times spent on delineation\cite{balagopal2021psa, zhang2023segment}. Beyond auto-segmentation, AI-based algorithms were also implemented in tumor staging\cite{liang2022multi},  image registration\cite{cao2018deformable, ding2023deep}, automatic treatment planning\cite{wang2020review}, quality assurance\cite{vandewinckele2020overview},  outcomes prediction\cite{liu2017exploratory, wang2019expression, yang2022exploratory, yang2022empirical}, as well as other areas. 

The introduction of the transformer architecture marked a significant milestone in the development of deep learning models, leading to remarkable advancements in terms of parameter size and model complexity \cite{vaswani2017attention}. There has been an exponential growth in the scale of these models which were trained on massive amounts of text data. Notable examples of large language models (LLM) that have emerged include OpenAI’s generative pre-trained transformer (GPT), Google’s pathways language model (PaLM), etc. Some domain-specific LLMs can serve as virtual assistants, facilitating healthcare advice, medical decision support, and administrative tasks\cite{gilbert2023large} \cite{singhal2023large}. Building upon the success of pre-trained large models in natural language processing (NLP), researchers have embarked on exploring pre-trained large  models in the domain of computer vision. Large visual models (LVMs), such as vision transformers (ViT) \cite{wu2020visual} and VideoMAE V2 \cite{wang2023videomae}, exhibit exceptional accuracy in recognizing objects and scenes within images and videos. Researchers are exploring the potential of combining language and visual models to develop advanced AI systems that possess a more human-like understanding of the world \cite{pan2022contrastive}.  Since AGI models are trained on large and diverse datasets, AGI can achieve impressive zero-shot/few-shot generalization on unseen/limited datasets and perform various real-world tasks with proper prompts.

In this review, we will focus on the different aspects of radiation oncology and the emerging applications and potentials of AGI. The structure of this paper is as follows. In section 2, we introduced the application of LLMs in radiation oncology, including the automatic selection of radiotherapy modalities, patient follow-ups, knowledge extraction from multi-center data, and standardization of clinical data. Section 3 presents how the LVMs and multimodal models shape the domain of medical imaging, and the potentials in radiation oncology. Section 4 discusses the transformative potential of AGI in radiotherapy dose prediction and automatic treatment planning. Section 5 summarizes the deep learning methods in the generation of synthetic CT and talks about the potential of AGI models.  Section 6 focuses on the application of  deep neural networks in medical image registration.  In Section 7, we explore the future directions of AGI in radiation oncology. The potential developments and bottlenecks in the field are discussed as well. 

\section{Large Language Models for Radiation Oncology}

\begin{figure}
    \centering
    \includegraphics[width=0.8\textwidth]{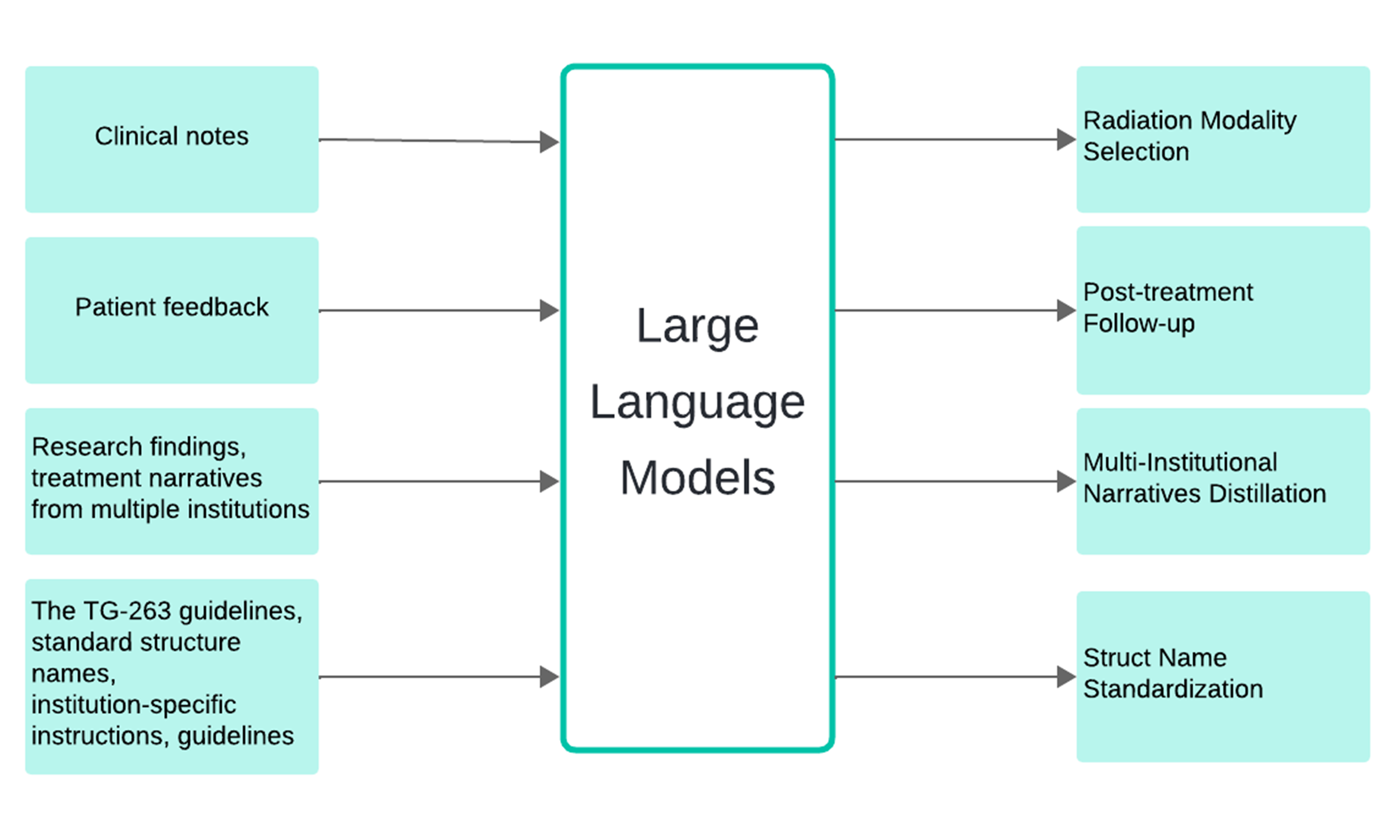}
    \caption{Large Language Models for Radiation Oncology}
    \label{fig:LLMs}
\end{figure}

Since the advent of the transformer architecture, there has been a remarkable burgeoning in the domain of Natural Language Processing (NLP). A plethora of sophisticated models have emerged, ranging from early pre-trained BERT models \cite{devlin2018bert,zhou2023comprehensive}, the Generative Pre-trained Transformer (GPT) series \cite{radford2018improving,radford2019language,brown2020language, ouyang2022training}, PaLM 2 \cite{anil2023palm} and recent open-source LLMs such as BLOOMZ \cite{muennighoff2022crosslingual} and the Llama series \cite{touvron2023llama, touvron2023llama2}. These architectures and domain-specific variants \cite{liu2023radiology,zhang2023biomedgpt,liu2023pharmacygpt,rezayi2022clinicalradiobert,ma2023impressiongpt} exhibit diverse capabilities in effectively tackling an extensive array of NLP tasks \cite{li2023artificial,guan2023cohortgpt,dai2023chataug,zhou2023fine,wu2023exploring} and downstream applications \cite{liu2023deid,liao2023differentiate,liu2023summary,liu2023evaluating,holmes2023evaluating}.

Radiation oncology is a dynamic field, marked by nuanced clinical decision-making and a relentless drive for efficacy and precision. Within this intricate landscape, the application of LLMs can usher in a new era of enhanced patient care and clinical efficiency. With a proven record of success in diverse science domains \cite{lu2023agi,mai2023opportunities,li2023artificial,latif2023artificial,zhao2023brain}, these models can sift through vast textual datasets to offer informed recommendations tailored to radiation oncology. In a seminary paper by Holmes et al. \cite{holmes2023evaluating}, researchers explored the capabilities of LLMs in radiation oncology physics. The study benchmarked four LLMs against medical physicists and non-experts using an exam developed at Mayo Clinic. ChatGPT (GPT-4) outperformed other models and even medical physicists in some tests, highlighting the potential of LLMs in specialized fields like radiation oncology physics. The impressive accuracy of GPT-4, and to a lesser extent GPT-3.5, in answering questions on the topic of radiation oncology physics suggests that LLMs may be adequate for a wide range of applications in radiation oncology. 

Figure \ref{fig:LLMs} illustrates the central role of LLMs in connecting data to clinical applications. For example, selecting the most appropriate radiation modality for each patient is a complex and time-consuming process in radiation oncology \cite{huynh2020artificial}. Presently, oncologists often have to spend considerable time reviewing extensive clinical notes, which include a plethora of text documents like demographics, operative notes, pathology and radiology reports, lab results, and discharge and consults notes. These notes are frequently peppered with subtexts and templated information from the Electronic Medical Record (EMR) systems, making them cumbersome to navigate. The manual nature of this task not only requires significant labor but is also prone to human errors and inconsistencies in judgment. Advanced LLMs, fine-tuned specifically on radiation oncology datasets, can ameliorate this issue. Such models can rapidly parse through intricate clinical narratives to recommend the most suitable radiation modality, thereby enhancing both efficiency and the likelihood of optimal patient outcomes.

LLMs also can be pivotal in post-treatment scenarios. By analyzing patient feedback in clinical notes or electronic communications, they can identify patterns suggestive of complications or side-effects. Furthermore, they can generate detailed, patient-specific educational materials, offering insights into the radiation procedures they underwent, potential side effects, and the logic behind the selected treatment modalities.

In addition, radiation oncology practices often differ across institutions. LLMs can process diverse clinical notes, research findings, and treatment narratives from multiple institutions \cite{liu2023summary,li2023artificial}. This vast reservoir of knowledge can then be used to inform best practices or highlight innovative treatment approaches that have found success in particular settings.

% by Jason Holmes
Another long-standing problem in radiation oncology is poorly labeled structure names \cite{cancers15030564, SCHULER2019191, SLEEMANIV2020103527, https://doi.org/10.1002/acm2.13662, healthcare8020120, biomedinformatics3030034}. When analyzing data for a set of patients where the structure contours are required, we want to identify a particular set of structure contours. To do this, the structure name is usually used, however there is typically a lot of variation in the naming of a structure across patients and institutions, which makes the subsequent data analysis very cumbersome. For example, for one patient, prostate might be "prostate" and for another it may be "pstate". For the femur head, the label may be "Femur\_head\_left" or "fem\_head\_l". It is very time-consuming and tedious to standardize the structure names and it is also prone to human errors.

This problem led to the development of a standard for labeling structure names. In 2018, the American Association of Physicists in Medicine (AAPM) created a task group (TG) report known as TG-263\cite{MAYO20181057}, which defined a standard for naming structures. However, even with a standard defined, clinics still often are slow to adopt the standard or choose not to. For clinics that choose to adopt the TG-263 standard, data prior to 2018 remains poorly labeled. Tons of these historical patient data in radiation oncology still need to be standardized if we want to take advantage of them for data-mining. For this reason, researchers have investigated methods for re-labeling structure names. 

Prospective methods typically employ search algorithms and look-up tables \cite{SCHULER2019191} or machine learning \cite{cancers15030564, SLEEMANIV2020103527, https://doi.org/10.1002/acm2.13662, healthcare8020120, biomedinformatics3030034}. The inputs for machine learning approaches always include the structure names and structures converted to binary masks. Additionally, inputs may include CT information, dose information, or reduced information such as the structure volume. Typically, the inclusion of additional information leads to better results, however studies consistently show that the most significant factor in the accuracy of these models to correctly re-label structure names is the structure name itself.

\begin{figure}
    \centering
    \includegraphics[width=0.6\textwidth]{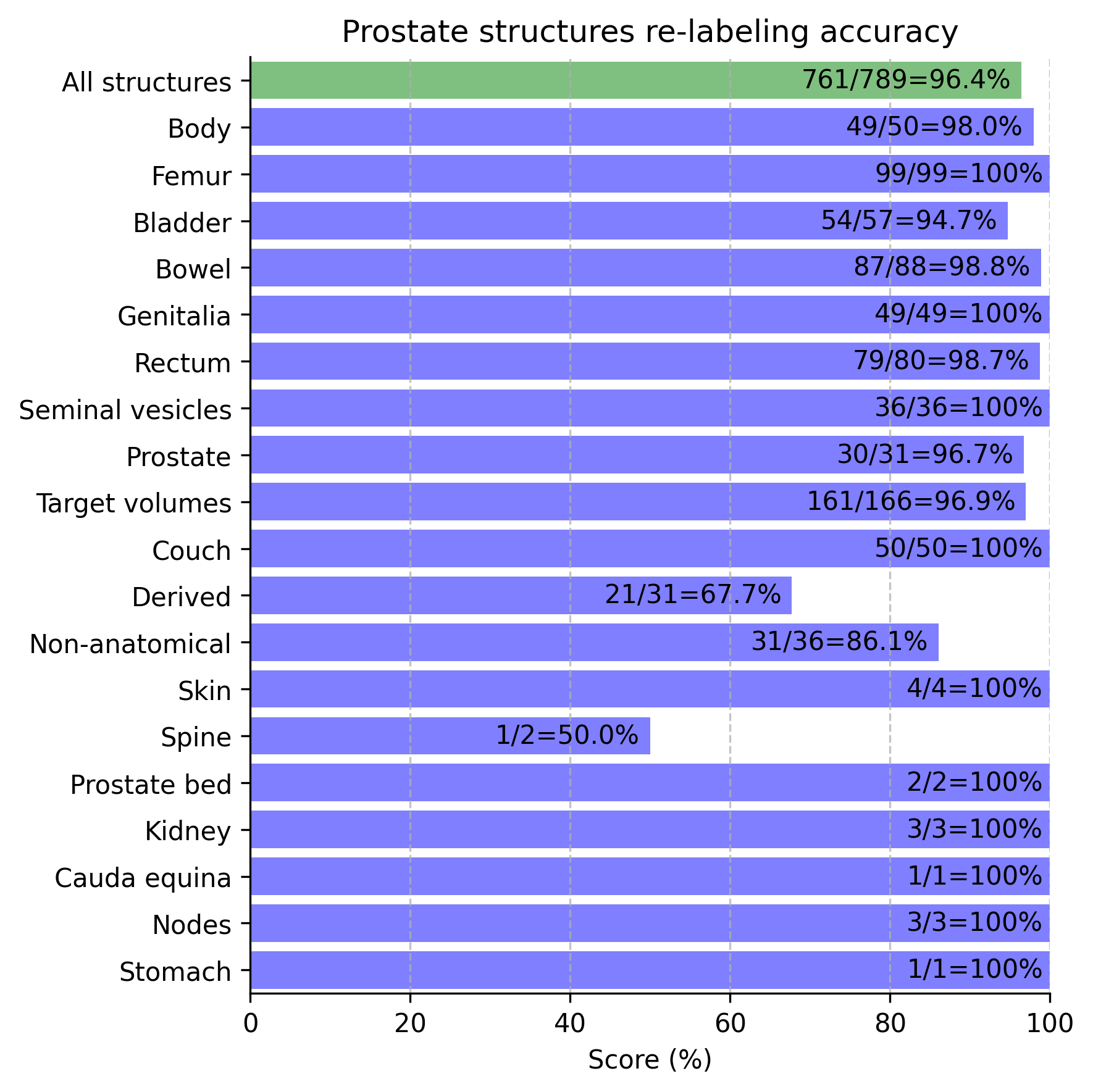}
    \caption{The accuracy of GPT-4 in re-labeling structure names according to the TG-263 report.}
    \label{fig:tg263}
\end{figure}

We have tested GPT-4's ability to re-label structure names on a per-patient basis for 50 prostate patients (never previously seen by GPT-4). The prompt includes the structure names, the TG-263 guidelines, and a list of standard structure names (no patient information). Importantly, the prompt also includes institution-specific instructions or guidelines. Figure \ref{fig:tg263} shows the results. These results are comparable to the prior reported studies, however there are important distinctions. In the prior reported studies, structure names are always treated as stand-alone, not as part of a set of structure names for the patient. In doing this, they lose important contextual information. Additionally, since GPT-4 does not need to be trained, we may include rare structure names where very little data exists. All of the prior reported studies only test the performance for a small set of structure names. The generality of GPT-4 is far beyond existing AI-based methods and may allow for easy implementation across institutions.

In the evolving landscape of radiation oncology, LLMs serve as valuable tools, amplifying the depth and breadth of human expertise. They promise a synthesis of vast knowledge, aiding clinicians in their quest to offer unparalleled patient care. However, it remains imperative that these models work in tandem with human experts, ensuring a blend of computational efficiency with compassionate care.

\section{Large Multimodal Models for Radiation Oncology}

%\subsection{Large visual models}
The success \cite{zhou2023comprehensive,liao2023mask,zhao2023generic,rezayi2023exploring,cai2023exploring,liu2023context,chang2023meta,zhong2023chatabl} of language models in the field of NLP also provides a revolutionizing paradigm for the advancement of the visual domain. The transformer architecture, in particular, has become building blocks for constructing LVMs \cite{kirillov2023segment,zhang2022dino,oquab2023dinov2}. Notable instances include the Vision Transformer (ViT) \cite{dosovitskiy2020image}, Swin Transformer \cite{liu2021swin}, VideoMAE V2 \cite{wang2023videomae}, and others \cite{xiao2023instruction,yu2023core,ma2022rectify,yu2022disentangling}. These LVMs undergo pre-training using extensive image datasets, equipping them with the capability to capture the complexities of image content and extract intricate semantic information. This empowers them to be highly effective across a wide range of downstream applications \cite{dai2023samaug,zhang2023segment,li2023artificial,dai2023hierarchical,zhang2023beam,bi2023community,zhang2023differentiating,ding2023deep,ding2023deep,ding2022accurate,liu2022discovering,dai2022graph}.

Recently, the emergence of the Segment Anything Model (SAM) \cite{kirillov2023segment} has introduced a novel approach to tackling downstream tasks. Unlike traditional pre-trained visual models, SAM operates as a promptable image segmentation model\cite{zhang2023segment}. It utilizes prompts provided by users to guide the model in accurately segmenting the desired areas. By skillfully engineering prompts, SAM is capable of addressing a broad range of downstream tasks effectively. Its remarkable zero-shot generalization capability underscores the pivotal role of prompt engineering in enhancing the efficacy of downstream tasks.

% \subsection{Large multimodal foundation models}

\begin{figure}
    \centering
    \includegraphics[width=1.0\textwidth]{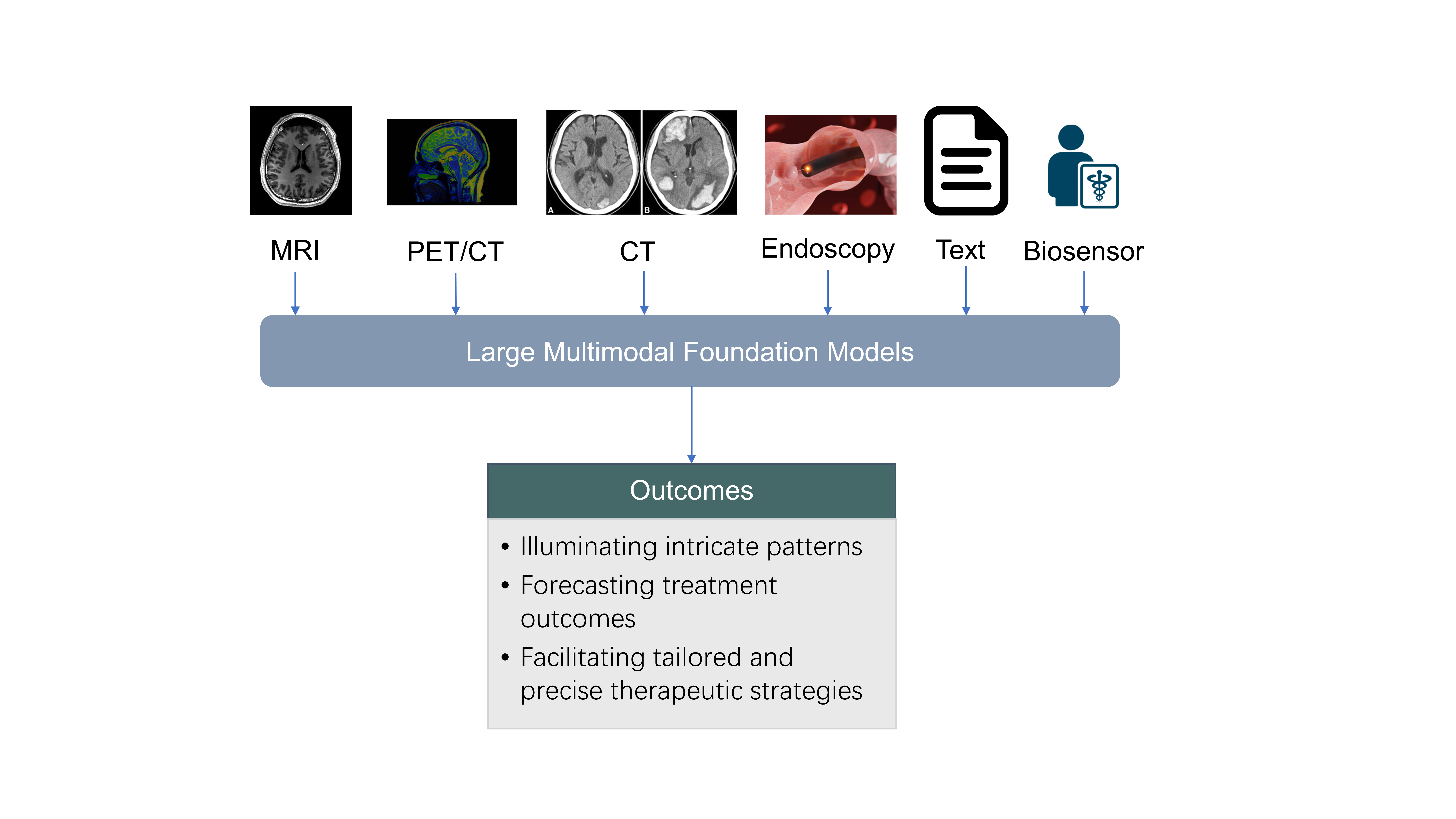}
    \caption{Large Multimodal Foundation Models for Radiation Oncology}
    \label{fig:Large Multimodal Foundation Models}
\end{figure}

The progress made in the field of LLMs and LVMs has provided a robust groundwork for the evolution of large multimodal foundation models \cite{fei2022towards}, marking a significant milestone in the integration of language and visual modalities. This harmonious combination of various modalities holds the promise of unraveling complex real-world situations and enhancing modeling capabilities. Capitalizing on this complementary synergy, the convergence of unimodal large language models and vision models has gained significant attention, ushering in the era of large multimodal models with boundless potential.

Transitioning from an unimodal framework to a multimodal one requires adjustments in both data and model aspects. In the realm of data, researchers commonly pivot towards creating multimodal-instruction text datasets. This can be achieved by either reconfiguring existing benchmark datasets \cite{liu2023visual, wang2023visionllm, chen2023x, xu2022multiinstruct} or employing innovative self-instruction methodologies \cite{zhu2023minigpt, liu2023visual, zhao2023chatbridge}. On the model horizon, a prevailing approach entails the fusion of information from disparate modalities into LLMs, endowing them with robust reasoning capabilities. Existing studies have taken one of two paths: the direct alignment of multimodal embeddings with LLMs \cite{liu2023visual,liu2023visual, wang2023visionllm, chen2023x} or the utilization of expert models to translate data from various modalities into a form assimilable by LLMs \cite{li2023videochat, yang2023gpt4tools}. Through these methodologies, these works reshape LLMs into multimodal conversational agents \cite{zhu2023minigpt, liu2023visual, li2023videochat} and versatile task solvers \cite{liu2023visual, wang2023visionllm, xu2022multiinstruct} by fine-tuning their performance based on multimodal instructions. 

Across various domains, large multimodal foundation models hold profound promise and unparalleled potential. Radiation Oncology \cite{holmes2023evaluating} is such a quintessential example of a multimodal domain, where multimodality takes on a significant role. Multimodality in radiation oncology refers to the convergence and effective utilization of diverse medical imaging data and complementary information sources. This integration is aimed at elevating the precision and efficacy of radiation therapy. The significance of large multimodal foundation models is notably exemplified in this domain. These models possess the capacity to illuminate intricate patterns, forecast treatment outcomes, and facilitate tailored and precise therapeutic strategies. Their potential is not confined to augmenting existing processes; instead, they hold the power to catalyze a transformative shift within the field. By charting a course towards data-driven insights, these models lay the foundation for a new era of data-driven radiation oncology. Figure \ref{fig:Large Multimodal Foundation Models} highlights some applications of multimodal foundation models in radiation oncology. 

One pivotal part of radiation therapy workflow is the accurate prediction of patient outcomes, which include metrics like tumor control, toxicity levels, and overall survival rates \cite{lambin2013predicting}. Patient outcome modeling plays a vital role in personalized cancer management across the spectrum of available treatment modalities and has specific applications in areas such as adaptive radiotherapy \cite{feng2022gpu}\cite{cui2022artificial}. Conventionally, this process demands the integration of diverse data types, including clinical, imaging, treatment, dosimetric, and biological data. The unstructured nature of clinical notes adds to the complexity, as it obliges clinicians to dedicate extensive hours to data extraction and analysis. Domain-adapted AGI models can dramatically transform this workflow by swiftly navigating through the clinical notes to extract relevant information. In doing so, these models contribute to the development of more robust predictive algorithms. Automation in this domain not only streamlines operations but also holds the potential to increase the accuracy of outcome predictions, thus informing better treatment planning and elevating the standard of patient care.  

In recent years, AI-based outcome models have made significant progress, such as acute skin toxicity for breast cancer\cite{saednia2020quantitative}, radiation pneumonitis for lung cancer\cite{cui2021integrating}, and overall survival for liver cancer,\cite{wei2021deep} etc. However, the successful integration of outcome models into clinical practice is not solely contingent upon the accuracy of the models themselves. It also hinges upon the ability of clinicians to interpret and comprehend the specific decisions made by the models. Consequently, the interpretability and explainability of the AI or AGI models hold equal importance to their accuracy when considering their clinical implementation.

\section{AGI for Treatment Planning}

Radiation therapy serves as a key treatment strategy in clinical practice. Over recent years, its effectiveness has been remarkably enhanced, largely attributed to state-of-the-art modalities like Intensity-Modulated Radiation Therapy (IMRT) and Intensity-Modulated Proton Therapy (IMPT) \cite{bortfeld2006imrt, bucci2005advances, liu2012robust, liu2016exploratory, liu2012influence, liu2013ptv}. This advancement has dual implications: It has improved treatment plan quality but has concurrently augmented the intricacy of these plans with longer planning duration and potential challenges in maintaining treatment accuracy. Clinicians often have to balance lots of planning-related parameters which necessitates intricate communications between dosimetrists and physicians to fine-tune the planning parameters. As a remedy to this, the research landscape is increasingly focusing on optimizing dose distribution and setting reasonable constraints\cite{wang2020review, nguyen2022advances, matney2013effects, liu2018impact}. The overarching ambition is to efficiently and accurately generate an optimal dose distribution comparable to a manual plan based on previous treatment planning knowledge and ensure both time efficiency and uniform excellence, irrespective of the planner’s level of expertise \cite{an2017robust, liu2020robust}. 

In clinical practice, one common strategy for enhancing the efficiency and quality of manual treatment planning involves reviewing prior cases deemed to be exemplary. In particular, parameters from these previous cases—such as beam configurations and dose-volume histogram (DVH) objectives used in inverse planning—can be directly integrated into the current planning process or serve as benchmarks for decision-making \cite{green2019practical, wu2011adaptive}. Building on this concept, researchers have employed statistical models to distill specific attributes from these superior cases, utilizing the best of clinical acumen and knowledge. The initial class of methodologies focuses on predicting viable plan parameters, such as DVH objectives to guide the optimization process. Approaches informed by artificial intelligence (AI) display remarkable advantages over conventional rule-based algorithms. Commonly known as Knowledge-Based Planning (KBP), these techniques anticipate possible dose distribution patterns for a new plan by drawing insights from historical, high-quality plans. While atlas-based KBP methods identify the most analogous patients in the plan database to ascertain the optimal starting parameters for inverse planning, model-based KBP methods deploy various computational models trained on previous plans to predict ideal parameters for the new case \cite{haseai2020similar, mcintosh2015contextual, sheng2015atlas}. The parameters generated by KBP can streamline the optimization process, cutting down on the number of trial-and-error adjustments, as they are inherently more aligned with ideal outcomes than those derived solely from planners' experience. When it comes to describing the geometric relationship between organs-at-risk (OAR) and the planning target volumes (PTV), various metrics such as the overlap volume histogram (OVH), distance-to-target histogram (DTH), and out-of-field volume are commonly employed. These metrics can inform DVH predictions through machine learning techniques, including support vector regression (SVR) \cite{zhu2011planning, boutilier2015models, zhang2018ensemble}. One commercial embodiment of this DVH-KBP model is Varian's RapidPlan \cite{ma2017assessment}. Contrary to DVH prediction alone, 3D-dose prediction preserves spatial specificity. Given that the DVH serves as a one-dimensional representation of the three-dimensional dose distribution, it inherently loses some spatial information during the translation. Consequently, identical DVH curves could originate from divergent dose distributions, potentially resulting in clinically meaningful disparities. Physicians typically utilize both DVH and raw dose data to assess the quality of treatment plans. Therefore, dose prediction could offer certain advantages that DVH prediction alone may not capture. Currently, deep learning algorithms have been proven to achieve expert-level 3D-dose prediction accuracy but with enhanced efficiency, even in advanced therapeutic modalities like helical tomotherapy and proton radiotherapy \cite{fan2019automatic, liu2019deep, nguyen20193d, kandalan2020dose, zhang2023beam}. Figure 4 shows the workflow of deep learning-based dose prediction and clinical deployment for pencil beam scanning proton therapy \cite{zhang2023beam}. Moreover, predicted spatial dose distribution can be operationalized to generate clinically executable plans via dose-mimicking algorithms \cite{kierkels2019automated, meyer2021automation}.

\begin{figure}
    \centering
    \includegraphics[width=1.0\textwidth]{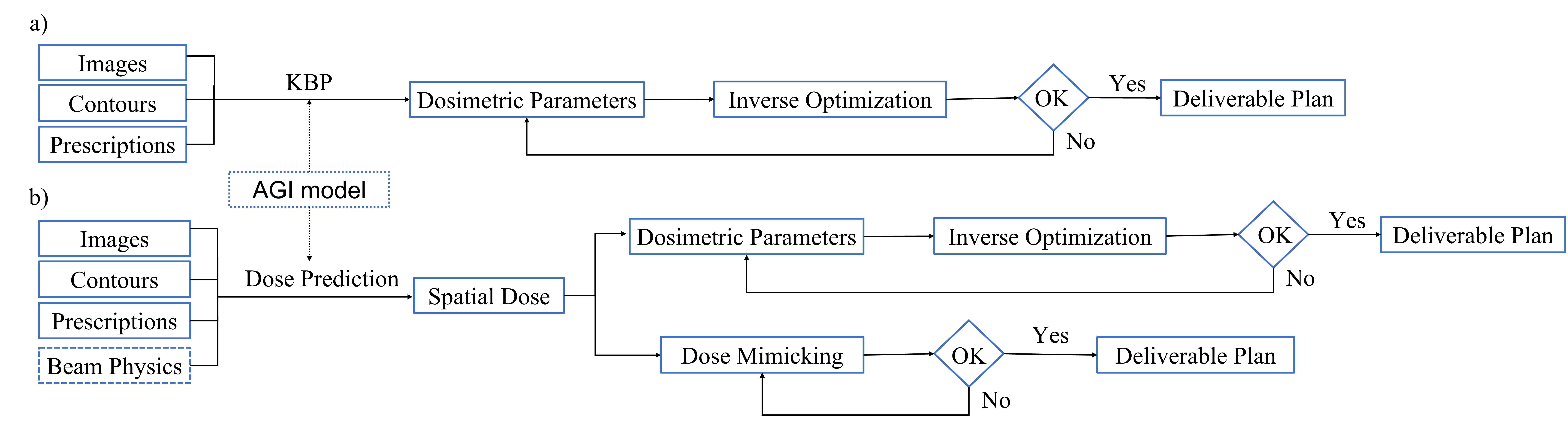}
    \caption{AI based treatment planning workflow. a) KBP method b) Dose Prediction method. The dotted box means an improved enhancement for the workflow.}
    \label{fig:AGI framework}
\end{figure}

In recent years, large foundational models such as ChatGPT and SAM have garnered substantial attention. As the burgeoning frontrunners in the AGI arena, these models offer tremendous promise for revolutionizing dose prediction in radiation therapy \cite{brown2020language, kirillov2023segment}. Current dose prediction paradigms, whether employing convolutional neural networks (CNN) or traditional machine-learning algorithms, often specialize in either particular types of clinical cases or specific anatomical locations \cite{sahiner2019deep}. This specialization complicates their deployment and broad applicability in clinical environments. In contrast, AGI frameworks possess an innate versatility, enabling a singular model to tackle a multitude of tasks \cite{goertzel2014artificial}. This capability could facilitate the development of a unified AGI-based model for comprehensive dose prediction, substantially reducing the complexity of its clinical integration. Moreover, AGI exhibits robust cross-modality learning potential; for example, SAM can transfer the segmentation knowledge acquired from natural images to medical image analyses. A recent study on SAM in clinical radiation therapy shows that the natural images trained SAM can achieve clinically acceptable Dice score >0.7 for most OARs segmentation across four disease sites, demonstrating super generalization capabilities across different disease sites and different modalities that make it feasible to develop a generic auto-segmentation model in radiotherapy with SAM \cite{zhang2023segment}. Given the often limited availability of training data for specific clinical scenarios or innovative techniques in clinical practice, AGI offers a robust solution for learning from a broad variety of knowledge. It permits fine-tuning based on a minimal dataset, generating reliable dose predictions. This is especially beneficial in cases of rare conditions or specialized techniques, such as MRI-based dose prediction, underscoring AGI's transformative potential in radiation therapy treatment planning \cite{pollard2017future}.

\section{Synthetic CT Generation Using Deep Learning}
Computed tomography (CT) is the primary imaging modality in the current practice of radiation therapy. It provides three-dimensional structural information of the patient for treatment planning, enabling electron density calibration required for dose calculation \cite{seco2006assessing}. Synthetic CT offers the possibility of reducing the additional dose of CT scan and expanding the usage of other imaging modalities in adaptive radiotherapy, such as MR, CBCT, and MVCT \cite{liu2020cbct}\cite{liu2019mri}\cite{zhao2021mv}\cite{feng2022gpu, shan2022virtual}. Magnetic resonance imaging (MRI) simulation has superior soft-tissue contrast compared with CT and delivers no ionizing radiation, which plays a vital role in target and organs-at-risk delineation. To create a treatment plan, MRI images have to be registered onto CT scans for dose calculation. Traditional synthetic CT generation methods can be grouped into three categories: bulk density override, atlas-based, and voxel-based methods \cite{JOHNSTONE2018199}. In recent years, many deep learning approaches have been proposed to predict the synthetic CT for MR-only radiotherapy \cite{cusumano2020deep}\cite{maspero2020deep}\cite{bird2021multicentre}\cite{lenkowicz2022deep}\cite{shafai2019dose}\cite{maspero2018dose}. There are mainly two categories, generator-only model, and generator and discriminator model. Han first proposed a deep convolutional neural network (DCNN) to model the MR-to-CT mapping using eighteen brain tumor patients \cite{han2017mr}. Other generator-only architectures, such as U-net  \cite{liu2019mr}\cite{liu2020abdominal}, Res-Net\cite{bahrami2021comparison}, multiple deep CNNs \cite{spadea2019deep}, deep embedding CNN \cite{xiang2018deep}, patch-based CNN \cite{dinkla2019dosimetric}, also have been used to estimate the electron density maps. The application of these methods extended from brain cancer \cite{han2017mr} to head and neck cancer \cite{dinkla2019dosimetric}, and prostate cancer \cite{xiang2018deep}.

Compared with the generator-only model, generative adversarial network (GAN) architecture involves two sub-models: a generator model to generate plausible data, and a discriminator model to determine fake from real data. Most researchers used the co-registered/paired MR and CT images as the training data of GAN and its variant models, such as GAN \cite{tang2021dosimetric}, conditional GAN \cite{maspero2018dose}, residual transformer conditional GAN \cite{zhao2023ct}, compensation cycle GAN \cite{zhao2023compensation}. To overcome the limitation of precisely aligned MR and CT images, some GAN models were developed to synthesize images from weakly paired, or unpaired data \cite{kang2021synthetic}\cite{abu2021paired}\cite{zeng2019hybrid}\cite{wang2023dc}\cite{brou2021improving}. Cycle-consistent GAN model was used in the training of weakly paired CT and MR images \cite{kang2021synthetic}. Zeng explored a hybrid GAN consisting of a 3D generator network and a 2D discriminator network, in which 3D generator was believed to model the 3D spatial information across slices \cite{zeng2019hybrid}. Augmented cycle GAN was proposed to generate synthetic CT using unpaired data from multiple scanners in different centers \cite{brou2021improving}. The model provided improved generalization performance and produced clinically acceptable synthetic CTs \cite{brou2021improving}. Although the robustness and generalizability of the generation models were considered in recent studies \cite{olin2021robustness}\cite{hsu2022synthetic}, the question of these techniques in clinical application is how to evaluate the quality of synthetic CT without acquiring the ground truth CT \cite{johnstone2018systematic}.  As far as we know, there are no studies using AGI to generate synthetic CT. The large vision models may be trained using diverse datasets from multiple centers, disease sites, and patient populations, which may ensure the model's robustness and generalizability.

\begin{figure}
    \centering
    \includegraphics[width=1.0\textwidth]{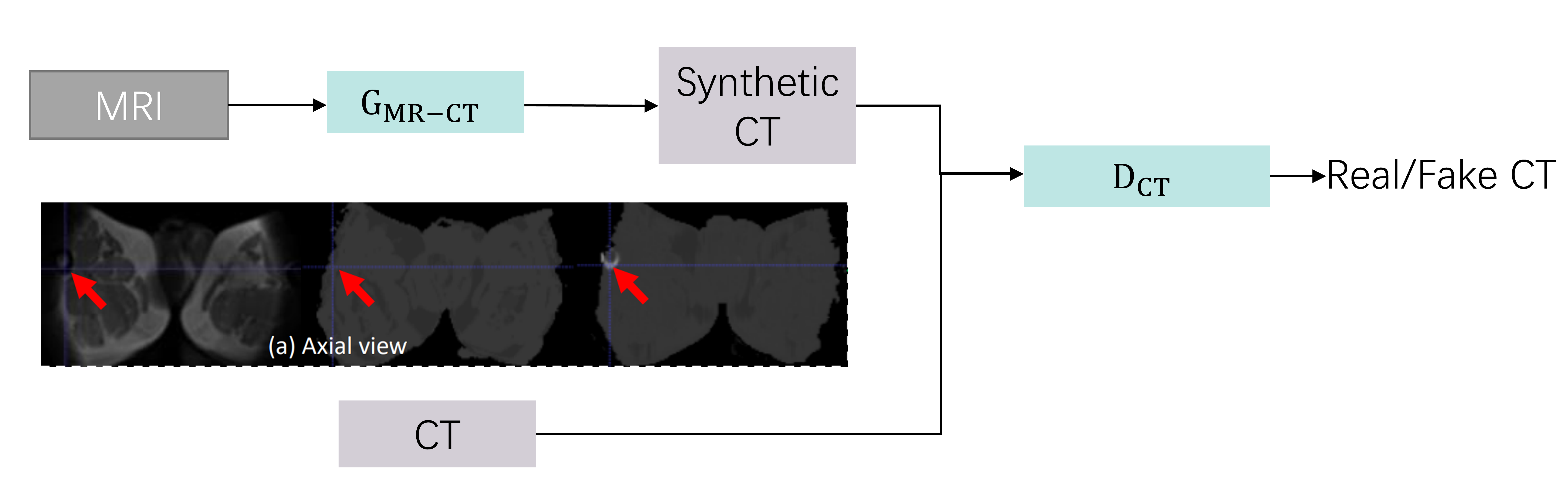}
    \caption{Synthetic CT generation using deep learning}
    \label{fig:Synthetic CT}
\end{figure}

\section{AGI for Image Registration}

Image registration, which aims to find the spatial relationship between two or multiple sets of images, is usually formalized as the optimization of a function balancing the similarity between images in terms of intensity, topology, or both \cite{oh2017deformable}. Compared to rigid image registration (RIR), the deformable image registration (DIR), which attempts to find the voxel-specific spatial relationship between two or multiple sets of images, has far more flexibility than RIR, thus, it can be used in more complicated clinical scenarios such as images with large anatomical structure changes. DIR has been extensively used in radiation therapy such as automatic segmentation \cite{thor2011deformable,hautvast2006automatic}, mathematical modeling \cite{sohn2005modelling, nguyen2009adapting, budiarto2011population, oh2014novel}, functional imaging \cite{guerrero2005quantification, yaremko2007reduction, yamamoto2011impact}, and dose deformation \cite{qi2015near, sharma2012dose, velec2011effect, yan1997adaptive, schaly2004tracking, christensen2001image, zhang2018impact}. Over the years, many conventional DIR approaches have been developed and adopted clinically. The conventional DIR approaches can be broadly categorized into two categories: parametric \cite{budiarto2011population, oh2014novel, yan1999model} and non-parametric \cite{vercauteren2007non, zhong2012finite, gu2013contour, nithiananthan2012extra} models. The parametric model generates deformable vector fields (DVFs) as a linear combination of its basic functions. The B-spline model is an example of such parametric models and it can handle the local change of a voxel by linear regression from nearby voxels within a certain distance. This property significantly reduces the computation time and memory required. Yet, the results can only be used for CTs with some strict conditions, such as breath-holding or respiratory gating, which limits its wide applications in clinics. In contrast, non-parametric models such as demons-based methods calculate transformation vectors of all voxels, thus achieving more accurate DVFs, but requiring more computation time and memory than the parametric models. 

Recently, several deep learning-based methods have been developed to speed up DIR in medical image analysis. According to the supervision used in model training, it can be broadly categorized as, supervised and unsupervised learning-based DIR approaches. For the supervised learning-based DIR methods, the ground-truth DVF is needed as the supervision. Yang et al. \cite{yang2016fast} proposed a two-steps deep learning framework for predicting the momentum parameterization for the large deformation diffeomorphic metric mapping (LDDMM) model. The proposed deep learning framework consists of two auto-encoder networks with the same architecture, in which the first auto-encoder is used to estimate the initial patch-wise momentum and the second one further tunes the initial patch-wise momentum. Although the proposed method is much faster comparing to the conventional DIR approaches, the computational complexity is higher than a typical single-step deep learning network. Besides, since it has two cascade networks, the symmetrical error may accumulate as the layers go deeper. The supervised learning-based DIR for other disease cites have also been studied \cite{cao2018deep, onieva2018diffeomorphic, lv2018respiratory}. Nevertheless, the generation of the ground-truth DVFs that are used for model training can be time-consuming as well, besides, the computed ground-truth DVFs may also be different from the real DVFs, thus introducing unexpected errors. Therefore, the unsupervised learning-based DIR approach which learns the similarity between the ground-truth image and wrapped image is more practical and favorable in clinical applications. Balakrishnan et al. \cite{balakrishnan2019voxelmorph} proposed a UNet-like model termed as VoxelMorph to learn the DVFs from pairs of magnetic resonance images (MRIs) (i.e., moving images and fixed images), then the generated DVFs and moving images go through a non-learnable spatial transformation to form the final generated warped images that resemble the fixed images. The VoxelMorph can achieve comparable performance as the state-of-the-art conventional DIR methods, whereas it is orders of magnitude faster. Thus, it has been widely used in medical image analysis. Most of these methods have been proposed for MRIs, which typically have high-resolution and rich anatomical information, whereas in radiation therapy the commonly used image modality is CT with a relatively low resolution. Recently, \cite{ding2023deep} have extended the unsupervised learning-based DIR to CT modality and yielded  fast and accurate results. 

With the advent of large multimodal foundation models, there has been a surge in the development of AGI-based image processing models, most notably represented by SAM, which has consequently accelerated research into the capabilities of AGI for image registration tasks. The general-purpose nature of AGI models makes them particularly well-suited for complex tasks like multi-modal image registration. Preliminary research reveals that SAM-enabled systems can perform real-time, accurate tracking and mapping of reference points in deformable images. For example, SAM has been shown to be effective in tracking respiratory motion within lung images. This capacity for continuous, point-to-point tracking has promising implications for its broader adoption in medical image registration scenarios \cite{rajivc2023segment}.

\begin{figure}
    \centering
    \includegraphics[width=0.6\textwidth]{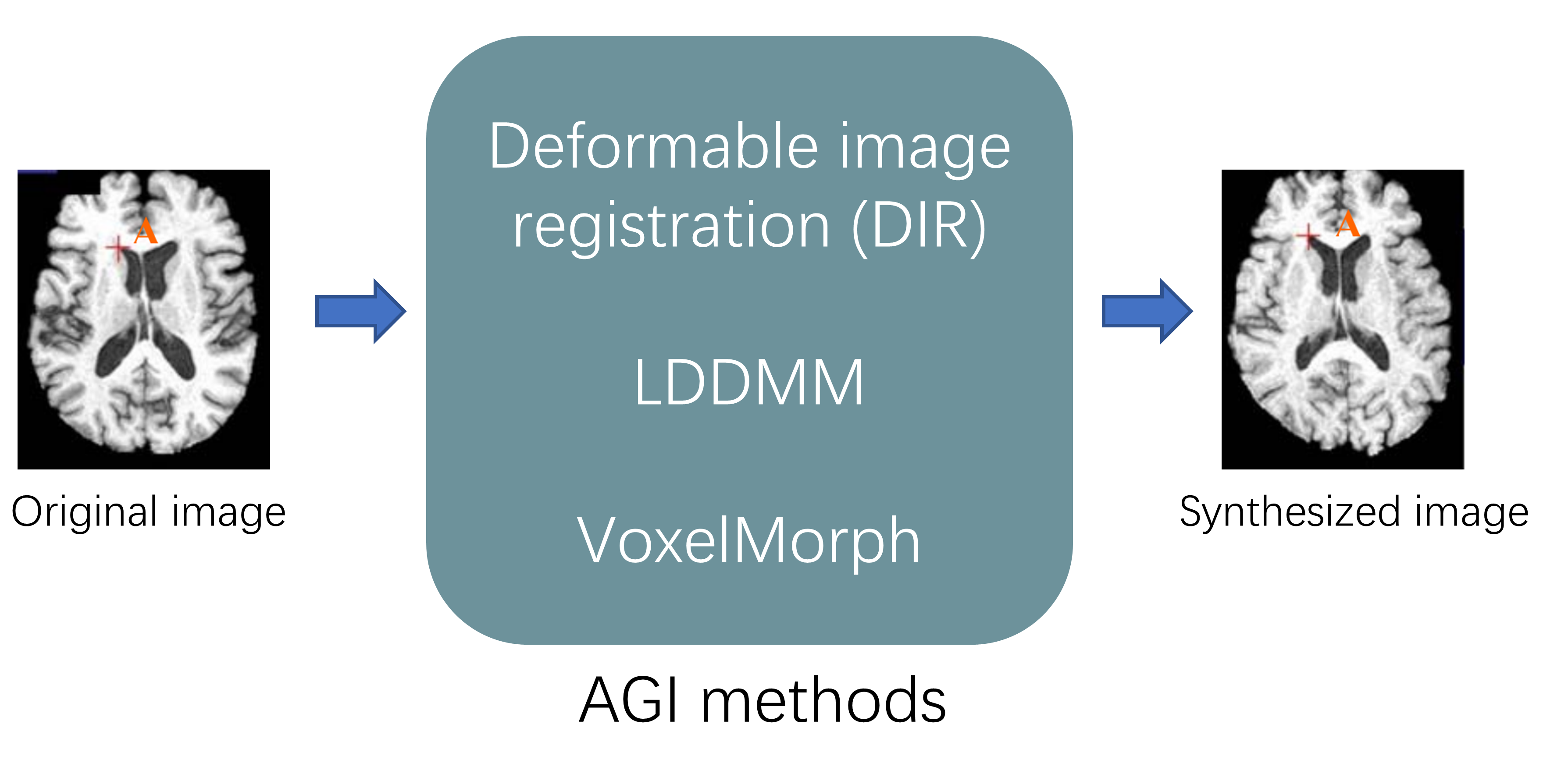}
    \caption{AGI for Image registration}
    \label{fig:Image registration}
\end{figure}

\section{Discussion and Conclusion}

In this review, we have outlined how AGI can revolutionize radiation oncology, enhancing healthcare standards. The key insight lies in AGI's capacity to leverage large-scale multimodal clinical data. Among the six stages of the radiotherapy workflow,  initial consultation, simulation, treatment planning, treatment delivery, treatment verification, and patient follow-up all involve visual and linguistic information. The processing and comprehension abilities of AGI in relation to visual and linguistic information have the potential to provide support throughout every stage of radiotherapy. This could result in improved radiotherapy safety and precision, enhanced efficiency, and favorable patient outcomes. However, the realization of AGI's potential in radiotherapy necessitates seamless integration with existing medical systems, and the performance of AGI is limited by its dependence on domain-specific knowledge. To surpass this limitation, future efforts should prioritize broadening the scope of knowledge by integrating diverse and comprehensive clinical datasets, utilizing interdisciplinary approaches, and encouraging interdisciplinary collaborations with clinical experts in radiation oncology.

As far as we know, LLMs, like GPT-3, are trained on text from multiple sources which include web pages, internet-based books, and Wikipedia. The LVMs are trained on the natural image datasets. Vision-language models are typically trained on multi-modal datasets harvested from the web in the form of matching image/video and text pairs. Although AGI generalized human cognitive abilities faced with unfamiliar tasks, most of the AGI models were not designed to provide high-quality clinical applications. To fine-tune these LLMs, LVMs, or vision-language models with high-quality medical data holds promising potential. By incorporating such data, AGI models can benefit from a broader range of clinical scenarios, leading to improved accuracy and performance in the applications of radiation oncology. 

One of the challenges is the data standardization. The inconsistency of structure names in radiotherapy poses a significant challenge when employing automated methods. As AAPM task group 263 reported, it is important to follow a standardized target and organs-at-risk naming rules for AGI training, data sharing between multi-centers, and quality assurance \cite{schuler2019big}. Data sharing presents an additional challenge in building a large, high-quality medical training dataset due to factors like privacy concerns, legal and ethical restrictions, and standardization obstacles to disseminating individual patient records. Data sharing has become increasingly prevalent in some fields of medical research, especially among genomics researchers and groups conducting systematic reviews and meta-analyses. Investigators still have concerns about sharing individual patient data from clinical trials \cite{ross2012importance}. Data interpretation in radiotherapy poses a notable challenge that warrants careful consideration to ensure meaningful analysis. This challenge arises from the inherent complexities of the clinical data, treatment uncertainties, and errors/variations in data recording. Integration of heterogeneous data in radiotherapy, including patient consultation, clinical examinations, medical imaging, treatment plans, verification records, machine logs, and patient outcomes, presents difficulties due to variations in formats, resolutions, etc.

Interdisciplinary collaboration teams play a vital role in ensuring the clinical applications of AGI models. These teams offer diverse perspectives and expertise, enabling comprehensive feedback throughout the development and validation stages. Involving end-users, such as radiation oncologists, and medical physicists, in the development process through user-centered design methodologies\cite{li2023artificial}. It ensures that the AGI models meet the domain-specific needs and preferences, enhancing the usability and acceptance of the models. The involvement of professionals from diverse disciplines ensures a comprehensive approach to address precision, ethical considerations, and regulatory compliance. In addition, developing AGI models that collaborate with medical professionals, rather than replacing them, offers several advantages. This approach allows for the maximization of AGI benefits while avoiding the resistance of medical professionals and optimizing their workload. With the development of AI, radiotherapy staffs prioritize continued education using AI to preserve their skills, such as manual segmentation ability\cite{ip2021current}. The introduction of AGI models will shape the evolving role of medical professionals in the radiotherapy department. With AGI handling micro/macro processes, the focus of radiotherapy staffs' work will shift towards verifying the model performance quality. As a result, the continued education of radiotherapy staffs will need to adapt accordingly. 

AGI holds great promise for radiotherapy, offering advancements in patient consultation, image registration, structure segmentation, radiation dose prediction, auto-treatment planning, patient outcomes prediction, etc. However, challenges remain regarding clinical datasets, regulation, and interdisciplinary collaboration considerations. Despite these challenges, the opportunities presented by AGI in enhancing radiation oncology and improving clinical automation are significant. This review aims to serve as a reference and catalyst for further exploration in this rapidly evolving field, stimulating advancements and discussions for the benefit of healthcare. 

\bibliography{LLM_refs}
\bibliographystyle{unsrt}

\end{document}